# Long-Wavelength Optical Properties of a Plasmonic Crystal

Cheng-ping Huang[1,2], Xiao-gang Yin[1], Qian-jin Wang[1], Huang Huang[1], and Yong-yuan Zhu[1]

[1]*National Laboratory of Solid State Microstructures, Nanjing University*

*Nanjing 210093, P.R. China*

[2]*Department of Applied Physics, Nanjing University of Technology*

*Nanjing 210009, P.R. China*

Abstract

The long-wavelength optical properties of a plasmonic crystal composed of gold nanorod particles have been studied. Because of the strong coupling between the incident light and vibrations of free electrons, bulk polariton excitation can be induced and a polaritonic stop band will be created. An analytical model has been established, which was confirmed by the numerical simulations. The similarity between the plasmonic and ionic crystals, as well as various optical properties, may be of both fundamental and practical interests.

* Email: cphuang@njut.edu.cn, yyzhu@nju.edu.cn

Recently, plasmonic materials composed of nanostructured metals have been intensively explored in realizing enhanced optical transmission, optical magnetism, as well as negative refractive index [1-3]. Employing plasmonic photonic crystals to produce a photonic stop band also opens up a new way for manipulating the motion of photons. In a metal surface, surface-plasmon polariton (SPP) mode, originating from the interaction of light with the surface charges, can be supported. Stimulated by the concept of dielectric photonic crystal, a two-dimensional SPP crystal, a metal surface textured with periodic particles or nanoholes (the lattice period is comparable with the SPP wavelength), has been constructed [4, 5]. Near the Brillouin zone boundary, a stop band for the surface mode can be created due to the Bragg reflection. Such structure has been used to make subwavelength plasmonic waveguide, highly dispersive photonic elements, and other devices [6, 7]. In addition, because of the larger dielectric contrast between the metal and dielectric medium, metallic nanostructures are also favorable for designing three-dimensional plasmonic photonic crystal that operates in the optical frequencies [8-11].

The Bragg reflection may not be the sole mechanism for stop band formation [12]. In this Letter, we suggested that a strong coupling effect, between the incident light and vibrations of free electrons, may exist in a plasmonic crystal, i.e., a three-dimensional array of nanorod particles. Due to the coupling effect (rather than the Bragg reflection), bulk polariton excitation can be induced and a polaritonic stop band will be created. Compared with a common photonic crystal, here the lattice period can be deep subwavelength. We have established an analytical model, which was confirmed by the numerical simulations. It is interesting to notice that the effect suggested here is similar to that obtained in an ionic crystal [13, 14], where strong coupling between the photons and lattice vibration is present and phonon-polariton mode is generated. Thus our results also bridge the connection between the artificial (plasmonic) and real (ionic) crystals.

The plasmonic crystal under study is composed of gold nanorod particles, which are arranged in a simple-cubic lattice in a host medium. Figure 1(a) shows the schematic view of the structure. Here, the lattice constant is $d$, the permittivity of host medium is $\varepsilon_d$, and the nanorod has a length of $l$ and radius of $r_0$. Supposing that the radius of nanorod is smaller than the skin depth ($r_0 < \delta \sim 20 nm$), the fields inside the nanorod can be taken to be homogeneous. We also assume that the sizes of nanorod are much larger than the Fermi wavelength ($r_0 >> \lambda_F \sim 0.5 nm$) and that the mean level spacing or Kubo gap ($\Delta E = 4 E_F / 3N$, where $E_F$ is the Fermi energy and N is the number of electrons) is very small compared with the thermal energy $kT$ (For a very small metal particle with a larger Kubo gap, the quantum size effect will be dominant [15]). Consequently, the quantum effect in the nanorods can be ignored and a classic description of the effect is applicable.

When the incident light propagates with the electric field along the rod axis, the free electrons can be excited and positive and negative charges will accumulate on the opposite sides of the nanorods, thus forming the electric dipole moment. This dipole moment stemming from the electronic motion is equivalent to that caused by the relative motion of ions in an ionic crystal. On the other hand, the electric dipoles will emit the electromagnetic waves, which further interfere with the incident light. The above effect can be enhanced in certain conditions. As we know, due to the near-field coupling, a metallic particle will interact with its neighbors such that a particle plasmon wave (or polarization wave) of transverse or longitudinal nature can propagate along the array of particles [16]. The dispersion relation for the transverse particle plasmon wave, which is similar to that of optical branch phonon mode in an ionic crystal, will intersect the light line. Consequently, near the crossing of dispersion curves, the incident light couples strongly to the transverse plasmon or polarization wave of the rod lattice. In that case, the propagation mode is neither a pure photon mode nor a pure plasmon mode. It is called a polariton mode, a mixture of the photons and polarization waves [14].

The plasmon resonance in a gold nanorod (along the rod axis) can occur at a much longer wavelength [17]. Here, we are interested in the case that the lattice constant is very small compared with the light wavelength and thus, the photons interact strongly with the transverse plasmon wave near the center of Brillouin zone. Similar to an ionic crystal [13], the fundamental equations governing the coupling effect in a plasmonic crystal may be given, in the long wavelength approximation, as

$$\ddot{W} = b_{11}W + b_{12}E, \\ P = b_{21}W + b_{22}E. \quad (1)$$

Here, $W$ represents the motion of free electrons in the nanorod, $E$ is the electric field of the light, and $P$ is the dielectric polarization induced by the electronic motion and the electric field. And, $b_{11}, b_{12}, b_{21}$, and $b_{22}$ are four unknown coefficients.

To illustrate the above idea, we first consider the motion equation of free electrons in a nanorod [see Fig. 1(b)]. Note that, in the long wavelength limit, the electronic motions of all nanorods can be taken to be almost identical. This enables us to use one parameter to characterize the long transverse plasmon waves. Moreover, different from an array of nanorod pairs which has a magnetic response [2, 18], the effect of light magnetic field can be neglected in our structure, as there is only one nanorod in a unit cell. Under the action of electric field of light, the motion of free electrons obeys the Newton's equation $m\,d^2z/dt^2 = -E_T e - \gamma m\,dz/dt$. Here, $m$, $e$, and $\gamma$ are the mass, charge, and collision frequency of the free electrons respectively, $z$ is the electronic displacement relative to the equilibrium positions, and $E_T$ is the total electric field in the nanorod. Previously, we have shown that a gold nanorod can be modeled as an $LC$ circuit having a self-inductance $L = (\mu_0 l/2\pi)\ln(l/2r_0)$ and a capacitance $C = 5\pi\varepsilon_0\varepsilon_d r_0/2$, where the two end faces of the nanorod become a circular capacitor [19]. Thus, the total electric field can be expressed as $E_T = E_{eff}^{(1)} + E_L + E_C$, where $E_{eff}^{(1)}$ represents the effective electric field imposed on the nanorod (caused by light and nanorod polarization), $E_L = -(L/l)d^2q/dt^2$ is the induced electric field associated with the self-inductance ($q$ is the charge carried by the capacitor), and $E_C = -q/Cl$ is the electric field resulted from the circular capacitor. Noticing that $q = -nes\,z$ ($n$ is the electron density, $s = \pi r_0^2$ is the rod cross-sectional area), one obtain

$$(m+\alpha L)\ddot{z} = -(\alpha/C)z - \gamma m\dot{z} - E_{eff}^{(1)}e, \qquad (2)$$

where $\alpha = ne^2s/l$ is a coefficient dependent on the nanorod size and the electron density (for gold, a realistic value of $n = 5.90\times 10^{28} m^{-3}$ is used in the calculation).

It can be seen from Eq. (2) that the free electrons confined in the nanorod will behave as the forced harmonic oscillators, which are characterized by an effective restoring force $F = -k_{eff}z$ ($k_{eff} = \alpha/C$) and an increased effective mass $m_{eff}$ ($m_{eff} = m + \alpha L$). The restoring force is related to the nanorod capacitance, where the accumulated charges in the capacitor will prohibit the directional motion of electrons. The effective mass of electrons is increased due to the self-inductance of nanorod, an electromagnetic inertia of the system (the increased electron mass has also been suggested for the periodic metallic wires [20]). For a nanorod with the length 150nm and diameter 30nm, for example, the effective electron mass attains $m_{eff} = 1.38m$. According to Eq. (2), the resonance frequency of free electrons in a single nanorod is $\omega_o = (k_{eff}/m_{eff})^{1/2}$. Equation (2) also suggests that the optical response of the free electrons in a nanorod is similar to that of the bounded electrons in a classic atom. Thus the gold nanorods can also be taken as the plasmonic "atoms".

The dielectric polarization of the plasmonic crystal comes from both electronic displacement in the nanorods and polarization in the host medium. The motion of free electrons along the rod axis gives rise to a dipole moment $p = ql = -nels\,z$. In the long wavelength approximation, the macroscopic polarization due to the free electrons is $P_{rod} = -nels\,z/\Omega$, where $\Omega = d^3$ is the volume of unit cell. On the other hand, the macroscopic polarization of the host medium (related to the bounded electrons) can be written as $P_{host} = \varepsilon_0 \chi E_{eff}^{(2)}$, where $\chi$ is the molecule polarizability per unit volume and $E_{eff}^{(2)}$ is the effective electric field acting on the medium molecules. Hence, the total dielectric polarization of the crystal reads

$$P = -(nels/\Omega)z + \varepsilon_0 \chi E_{eff}^{(2)}. \tag{3}$$

To establish the relationship between Eq. (2), (3) and Eq. (1), the effective electric field imposed on the nanorods and medium molecules should be determined. According to Lorentz effective-field model, we have $E_{eff}^{(1)} = E + P_{rod}/3\varepsilon_0\varepsilon_d$ for the nanorod and $E_{eff}^{(2)} = E + P_{host}/3\varepsilon_0$ for the medium molecule. It is not difficult to find that, with the use of the effective field formula, Eq. (2) and (3) can be transformed to the following form:

$$\ddot{W} = -(\omega_o^2 - \frac{Q^2}{3\varepsilon_0\varepsilon_d M\Omega})W + \frac{Q}{(M\Omega)^{1/2}}E,$$
$$P = \frac{Q}{(M\Omega)^{1/2}}W + \varepsilon_0(\varepsilon_d - 1)E. \tag{4}$$

Here, $W = (M/\Omega)^{1/2}z$ is the motion parameter proportional to the electronic displacement, $M = nlsm_{eff}$ and $Q = -nlse$ are, respectively, the total effective mass and total charge of free electrons in a nanorod. In deriving Eqs. (4), Clausius-Mossotti formula $\chi = 3(\varepsilon_d - 1)/(\varepsilon_d + 2)$ has been used and the loss from the electronic collision in the nanorod has been neglected. One can see from Eqs. (1) and (4) that they have the same format, thus demonstrating the validity of original hypothesis. The unknown coefficients in Eqs. (1) are then determined, respectively, to be $b_{11} = -(\omega_o^2 - Q^2/3\varepsilon_0\varepsilon_d M\Omega)$, $b_{12} = b_{21} = Q/(M\Omega)^{1/2}$, and $b_{22} = \varepsilon_0(\varepsilon_d - 1)$.

Eqs. (4) becomes the basic equations that describe the coupling effect between the photons and the long transverse plasmon wave. With the use of Eqs. (4), the dielectric function can be obtained as

$$\varepsilon(\omega) = \varepsilon_d + \frac{fm/m_{eff}}{\omega_s^2 - \omega^2}\omega_p^2. \tag{5}$$

Here, $\omega_s = (-b_{11})^{1/2} = (\omega_o^2 - fm\omega_p^2/3m_{eff}\varepsilon_d)^{1/2}$ is the eigenfrequency of the system, which is smaller than that of a single gold nanorod ($\omega_o$) due to the collective effect; $\omega_p = (ne^2/m\varepsilon_0)^{1/2}$ is the bulk plasma frequency of the gold; and $f = ls/\Omega$ is the filling ratio of the nanorods. Moreover, the dispersion relation of the polariton mode can be deduced with Maxwell equations and Eq. (5), yielding

$$\frac{c^2 k^2}{\omega^2} = \varepsilon_d + \frac{fm/m_{eff}}{\omega_s^2 - \omega^2} \omega_p^2, \tag{6}$$

where $c$ is the light velocity in vacuum. The result resembles that obtained in an ionic crystal where the photons couple strongly with the transverse optical phonons and phonon-polariton dispersion is induced.

The polariton dispersion and dielectric function are plotted, respectively, in Fig. 2(a) and 2(b) as a function of normalized frequency ($\omega/\omega_s$). Without loss of generality, here the lattice constant is 80nm, the permittivity of host medium is 2.25, and the length and diameter of nanorod is 40nm and 10nm respectively [Note that, here, the Kubo gap ($\Delta E \sim 40\,\mu eV$) is three orders of magnitude smaller than the thermal energy at room temperature ($kT \sim 26\,meV$), thus justifying the classical approach]. This gives a resonance wavelength of $\lambda_s = 2\pi c/\omega_s = 960\,nm$, significantly larger than the lattice constant (the resonance wavelength of a single nanorod is $\lambda_o = 940\,nm$). In Fig. 2(a), the inclined solid line and the flat dotted line correspond, respectively, to the light wave and the long transverse plasmon wave without mode coupling. The region of crossover of the solid and dotted lines is the resonance region, where the photons couple strongly to the long transverse plasmon wave (the solid circles represent the coupled mode). Near the resonance the propagation mode is not a pure photon mode or a pure plasmon mode but a coupled wave field consisting of both components. The quantum of this coupled mode is called a polariton.

One important effect of the coupling is that a polaritonic stop band will be created in the frequency range $[\omega_s, \omega_t]$, where the dielectric function is negative [see Fig. 2(b)] and the wavevector becomes imaginary (thus the light propagation will be forbidden). Here, the upper cutoff frequency is determined, by setting $\varepsilon(\omega) = 0$, to be $\omega_t = (\omega_s^2 + fm\omega_p^2/m_{eff}\varepsilon_d)^{1/2}$. With the used parameters, the corresponding cutoff wavelength is $\lambda_t = 905nm$ (here the effective electron mass is $m_{eff} = 1.04m$). The relative band width, which is defined as the absolute band width divided by the eigenfrequency, is approximately $\eta = fm\omega_p^2 / 2m_{eff}\varepsilon_d\omega_s^2$. From it the relative band width is found to be $\eta = 6.3\%$, which is one order of magnitude larger than the filling ratio ($f=0.6\%$). In addition, when the loss from the free electrons is accounted [see Eq. (2)], a damping term $-i(m/m_{eff})\gamma\omega$ will appear in the denominator of dielectric function. This leads to a maximal imaginary part of dielectric function (see inset of Fig. 2) and a peak of absorption locating at the eigenfrequency. The absorption is equivalent to the infrared absorption in an ionic crystal.

To verify the polaritonic stop-band effect, we have calculated the transmission spectrum of a plasmonic crystal film analytically and compared it with the numerical results (the lattice parameters mentioned above are used). For a free-standing film with the thickness $h$, the transmission efficiency of a normally incident light is

$$t = \left| \frac{4k_0 k \exp(ikh)}{(k_0 + k)^2 - (k_0 - k)^2 \exp(2ikh)} \right|^2. \quad (7)$$

Here, $k_0$ is the wavevector in free space and k is the wavevector of polariton mode. The transmission spectrum calculated by Eq. (6) and (7) is presented as the solid circles in Fig. 3(a), showing a polaritonic stop band between 905 and 960nm (the film thickness is 1600nm, twenty unit cells thick). In the pass band, amplitude oscillation is observed due to the film Fabry-Perot resonance. As a comparison, Fig. 3(a) also presents the spectrum (the open circles) of the same structure simulated with the commercial software package FDTD Solutions 6.5 (Lumerical Solutions, Inc., Canada). In the numerical simulation, the gold is modeled with a lossless Drude model $\varepsilon_m = \varepsilon_\infty - \omega_p^2/\omega^2$, where $\varepsilon_\infty = 7$ and $\omega_p = 1.37 \times 10^{16}$ rad/s are used according to the experimental data of gold [21]. One can see that the numerical simulation agrees well with the analytical calculations, concerning the opening of stop band and amplitude oscillation in the pass band.

Additional insight can be provided by studying the transmission spectrum with a varying incident angle. Figure 3(b) presents the numerical results for TE polarization, where the electric field is fixed along the rod axis to maximize the coupling effect. The results show that the stop band formation is not dependent on the incident angle or following the Bragg diffraction. This can be understood, as the wavelength is much larger than the lattice constant and the light only "feels" an average response. In addition, we also calculated the transmission spectra, as shown in Fig. 3(c), of the structure considering the loss of gold (normal incidence). Here, the solid circles represent the analytical results of Eq. (7) and the open circles the numerical results using a lossy Drude dispersion (in both cases, $\gamma = 5 \times 10^{13}$ rad/s was used). A nearly perfect agreement between them is found. The results show that the stop band still survives but is enlarged due to the absorption.

In summary, the long-wavelength optical properties of a plasmonic crystal composed of nanorod particles have been studied. We emphasized the concept of polariton, which is due to the coupling between the photons and the long transverse plasmon wave. The polaritonic stop band, associated with the coupling effect rather than the Bragg reflection, has been suggested. The results also show that the long-wavelength method developed for an ionic crystal can be applied to a plasmonic crystal and that the artificial and classic lattices may share a common physics.

This work was supported by the National Natural Science Foundation of China (Grant No. 10874079, 10804051, and 10523001), by the State Key Program for Basic Research of China (Grant No. 2006CB921804 and 2010CB630703).

**Figure captions**

Fig. 1. Schematic view of the structure under study: (a) in the plasmonic crystal, gold nanorod particles are arranged in a simple-cubic lattice. The incident light propagates in the x-direction with the electric field along the rod axis; (b) the free electrons in the nanorod can be excited, leading to the accumulation of positive and negative charges on the opposite sides.

Fig. 2. Calculated polariton dispersion (a) and dielectric abnormality (b). Inset shows the complex dielectric function where the loss of gold is considered: the solid line and circles represent, respectively, the real and imaginary part of the function (here $\gamma = 5\times10^{13}$ rad/s has been used).

Fig. 3. (a) Analytically calculated (the solid circles) and numerically simulated (the open circles) transmission spectra (normal incidence and loss-free); (b) Simulated spectra (loss-free) with the incident angle being 0 (the solid circles), 15 (the open circles), and 25 (the open squares) degrees; (c) Spectra (normal incidence) accounting for the loss of gold: the solid and open circles represent the analytical and numerical results, respectively. Here, a free-standing plasmonic crystal film with the thickness 1600nm was used.

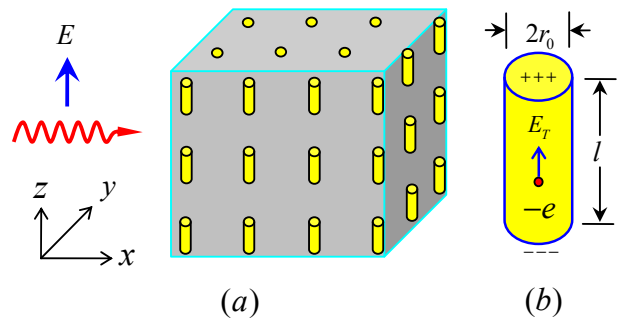

**Figure 1**

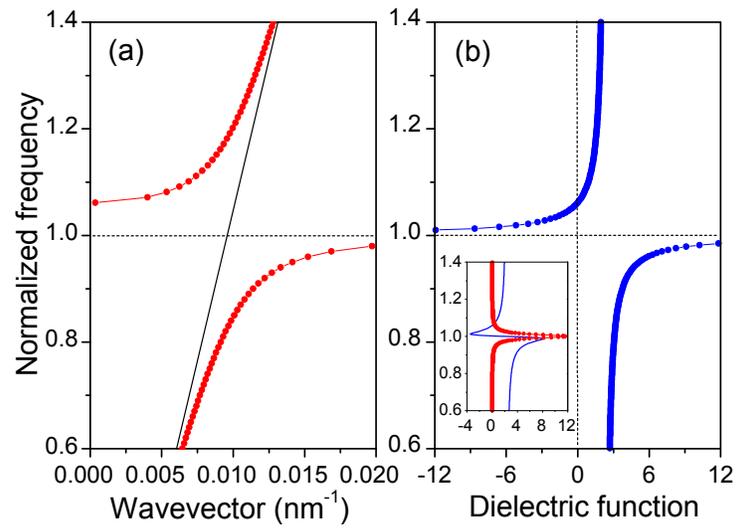

**Figure 2**

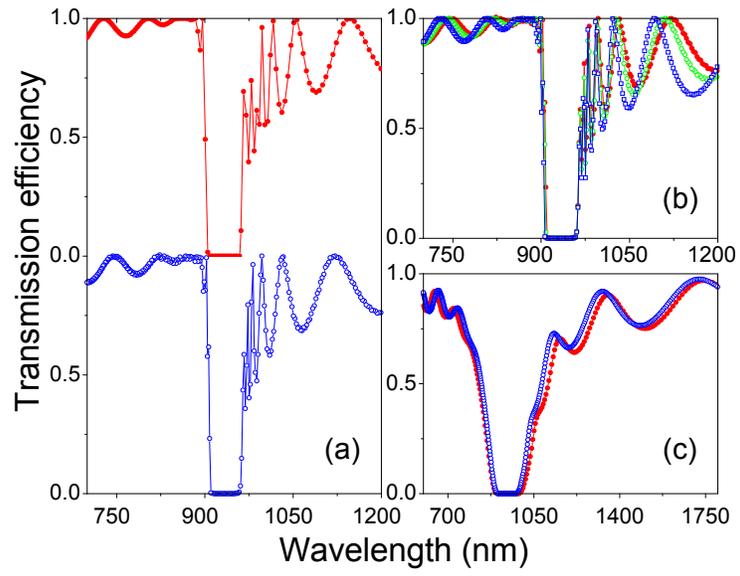

**Figure 3**